\documentclass[12pt]{amsart}
\usepackage{geometry} 
\usepackage{graphicx}
\usepackage{amssymb,latexsym,amsmath,mathrsfs} 

\title{Estimation and svm classification of glucose-insulin model parameters from OGTT data. An aid for diabetes diagnostics}

\author{Miguel Angel Moreles \and Joaquin Pe\~{n}a \and Paola Vargas \and Adriana Monroy
       \and  Silvestre Alavez}

\address{M. A. Moreles, J. Pe\~{n}a, P. Vargas \\
              Centro de Investigaci\'{o}n en Matem\'{a}ticas\\ Jalisco s/n, Valenciana\\
Guanajuato, GTO 36240,Mexico\\
              \email{moreles@cimat.mx}    \\     
                         A. Monroy \\
           Mexico General Hospital\\
Mexico City, Mexico \\
  S. Alavez \\
              Health Sciences Department \\
Metropolitan Autonomous University \\ 
Lerma, Mexico
}

\date{}

\begin{document}

\maketitle
\tableofcontents

\begin{abstract}
In the Oral Glucose Tolerance Test (OGTT), a patient, after an overnight fast ingests a load of glucose. Then measurements of glucose concentration are taken every 30 minutes during two hours. The test is used to aid diagnosis of diabetes, namely, type 2 diabetes mellitus and glucose intolerance. Several mathematical models have been introduced to describe the glucose-insulin system during an OGTT. Models consist on systems of differential equations where most parameters are unknown. Estimation of these parameters is an aim of this work.  In a minimal model, two of such parameters are proposed for classification by means of a SVM technique. Consequently, a case is made for this classification as an aid for diagnosis.
\end{abstract}

\section{Introduction}

According to the World Health Organization (WHO) global report on diabetes, people living with the disease has increased dramatically during the last four decades. Research on prevention, treatment, diagnostics, etc. is an urgent need. Our modest goal is to delve into the Oral Glucose Tolerance Test (OGTT) as an aid for diagnostics. 
 
More precisely, we are concerned with mathematical models of the glucose-insulin regulatory system during the OGTT. These models are in terms of Ordinary Differential Equations (ODE), and are short term in the sense that they describe the dynamics after an external source excites the system. In the OGTT, the external source is a load of glucose administered to the patient after an  overnight fast. Then glucose concentration is measured every 30 minutes. 

In practice, an ODE model is chosen and OGTT data is used to estimate parameters in the model.  This approach is of great interest, given the possibility to use the parameters of the models as indicators of diabetes development. 

Research along these lines is vast. A recent review of mathematical modeling of the glucose-insulin system is presented in  Palumbo et al \cite{palumboetal}. For a more computational, parameter estimation oriented review, see Rathee \& Nilam \cite{RatheeNilam}. These works provide an extensive bibliography, as well as the underlying biological mechanisms of the glucose-insulin system. We shall be concise on the latter.
 
A natural application of parameter estimation is to validate a model by data fitting. Our motivation is to take an step further and explore good fitting parameters for classification. In the context of the glucose-insulin system we start with actual data gathered at the Mexico General Hospital. In total, 80 female patients underwent the OGTT. 51 healthy, 4 with Impaired Fasting Glycaemia (IFG), 15 with Impaired Glucose Tolerance (IGT), 7 with both alterations and 3 with Diabetes Mellitus Type 2 (T2DM).

Models for the glucose-insulin system can be quite complex, involving several parameters in a system of ordinary differential equations. Consequently, second order minimal models, e.g. Caumo, Bergman \& Cobelli \cite{caumoetal}, are preferred. 

For a  first test of the methodology to be presented, we use the minimal model of Ackerman et al \cite{Ackermanetal}. Hereafter  referred as the basic model. This model is  probably the simplest. We find the obtained results highly satisfactory, which have led us to report our findings. 

The basic model can be reduced to a single differential equation corresponding to a harmonic oscillator. The state variable, is the the deviation from the stable fasting level of the glucose concentration. To solve the parameter estimation problem, we follow the bayesian approach.  The MAP estimator is chosen from the posterior. The handling of noisy data and assessment of the quality of point estimators, is straightforward in bayesian estimation. These features have made bayesian estimation a natural choice. A strong case is made in Pillonetto,  Sparacino \& Cobelli \cite{pillonettoetal}.

For each patient, four parameters are estimated. Amplitude and damping coefficient are used for classification. A Support Vector Machine technique is applied to the 80 pairs of parameters. The classification is correct for 85\% of the cases, and more importantly, the separating line appears as a transition line from healthy to mildly ill to diabetic. The application in mind is early detection of patients at risk.

Let us describe the content of this work. In Section 2 we discuss minimal models, introduce the basic model and its solution in the homogeneous damped oscillator case. Then the parameter identification problem of interest is posed. Main results are presented in Section 3. A sample of fitting curves for patients in various conditions and a SVM classification plot are shown. It is argued that one may build on this  SVM plot to develop an aid for early detection of pre diabetic patients. We leave for Section 4 the bayesian estimation. We focus on the damping coefficient, which was regarded unreliable in the past. In contrast, marginal posterior densities, show that the MAP estimator of this parameter is somewhat robust. We close with a section with conclusions and future work.

\section{The parameter estimation problem given OGTT data}
 
 \subsection{Minimal glucose-insulin models during an Oral Glucose Tolerance Test}
 
  The OGTT starts with an overnight fast. On arrival to the lab,  a glucose load of $75g$ is orally administered to the patient. Then glucose concentrations are measures at time $0$, $30,$ $60$, $90$, and $120$ minutes. 
  
Let $G(t)$ be the glucose concentration in the blood and $H(t)$ be the net concentration of a variety of hormones that influence the blood glucose levels.  For OGTT conditions, insulin is considered predominant and $H(t)$ is essentially its concentration.  
 
 Minimal models consist of second order ODE systems describing the kinetics of glucose concentration and insulin action, namely
 \begin{equation}
 \begin{array}{rcl}
 \frac{dG}{dt} & = & P(G,H,\mathbf{u})+ J(t) \\
  & & \\
   \frac{dH}{dt} & = & Q(G,H,\mathbf{u}),
 \end{array}
 \label{MinModel}
 \end{equation}
 Here $\mathbf{u}$ is the vector of parameters in the model. The glucose load is regarded as a function $J(t)$
 
 \subsection{The simplest model}
 
Following Ackerman et al (1969), it is assumed that after fasting, the patient's concentrations have stabilized to $G_0$ and $H_0$.  Consequently,
we have that 
 
 \[
 P(G_0,H_0,\mathbf{u})=0, \quad Q(G_0,H_0,\mathbf{u})=0.
 \]

 We study the small deviations
 \[
 g(t) = G(t)-G_0,\quad h(t) = h(t)-H_0.
 \]

Neglecting second order terms in Taylor's formula we are led to
\begin{eqnarray*}
   \dot{g} &= &-m_1 g - m_2 h + J \\ 
   \dot{h} &= &-m_3 h + m_4 g \label{eq6}
\end{eqnarray*}
where $m_1$, $m_2$, $m_3$, $m_4$, are nonnegative constants.

After some time $J(t)\equiv 0$. Thus, the system reduces to the following second order differential equation,

\begin{equation*} 
  \ddot{g} + 2\alpha\dot{g} + \omega_{0}^2g = 0
\end{equation*}

A viable interpretation of the glucose-insulin system, is that of a damped harmonic oscillator. Hence we assume
\[
\alpha^2 - \omega_{0}^2<0.
\]

It is readily seen that the general solution is

 \begin{equation}
     g(t) = Ae^{-\alpha t} \cos(\omega t - \delta).
  \label{GenSol}
 \end{equation}
for some constants $A$, $\alpha$, $\omega$, $\delta$.

\subsection{The parameter estimation problem}

Let us denote the OGTT data for patient $j$ by $G^j_0$, $G^j_{30}$,  $G^j_{60}$, $G^j_{90}$, $G^j_{120}$.

Glucose concentrations for 80 female patients have been provided with the following conditions:

\begin{itemize}
\item 51 Healthy (H). $j=1,\ldots,51$

\item 4 Impaired Fasting Glycaemia (IFG). $j=52,\ldots,55$

\item 15 Impaired Glucose Tolerance (IGT). $j=56,\ldots,70$

\item 7 with both alterations (IFG-IFT).  $j=71,\ldots,77$

\item 3 with Type II Diabetes Mellitus (T2DM). $j=78,\ldots,80$
\end{itemize}

\bigskip

Let us assume that $J(t)\equiv 0$ for $t> 0$ and define $g^j_t=G^j_t-G^j_0$ for $t=30,60,90,120$. 

The problem of concern is: Given OGTT data for patient $j$:  $g^j_{30}$,  $g^j_{60}$, $g^j_{90}$, $g^j_{120}$,  estimate the parameters
\[
\mathbf{u}_j=(A_j,\alpha_j,\omega_j,\delta_j)^t,
\]
constrained to

\begin{equation*}
     g^j(t) = A_je^{-\alpha_j t} \cos(\omega_j t - \delta_j).
\end{equation*}

\bigskip

\noindent\textbf{Remark. }(i) The problem is commonly posed as a constrained optimization problem. If $\mathbf{y}$ is the data vector, and $\mathcal{G}$ is the so called observation operator, we pose
\[
min\, \left\vert y-\mathcal{G}(\mathbf{u})\right\vert^2
\]
constrained to
 \begin{equation*}
 \begin{array}{rcl}
 \frac{dG}{dt} & = & P(G,H,\mathbf{u})+ J(t) \\
  & & \\
   \frac{dH}{dt} & = & Q(G,H,\mathbf{u}),
 \end{array}
\end{equation*}
In most cases, there is no analytical solution of this ODE system and a numerical method is required. \newline
(ii) We shall see below that the methodology to be introduced applies in general, and (i) is a particular case.

\section{Data fitting and SVM classification}  

\subsection{Curve fitting}
There are some well known methodologies for fitting a curve through data by means of point estimates. Our aim is to provide additionally a gauge of the robustness of the point estimators. By formulating the problem as one of bayesian estimation, this is accomplished through the posterior probability density function,  determined for each parameter.  Details below.

In this paragraph we just illustrate curve fitting using the MAP estimator for some patients in different diabetic conditions.  Healthy patients in Figure 1, whereas ill patients  in Figure 2.

 \begin{figure}[h]
\begin{center}$
\begin{array}{cc}
\includegraphics[width=2in]{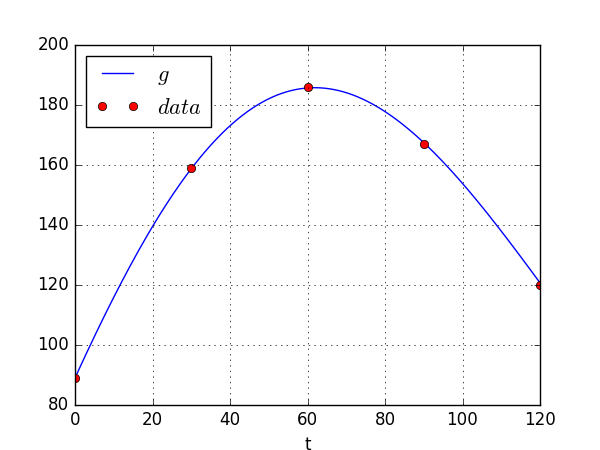}  & \includegraphics[width=2in]{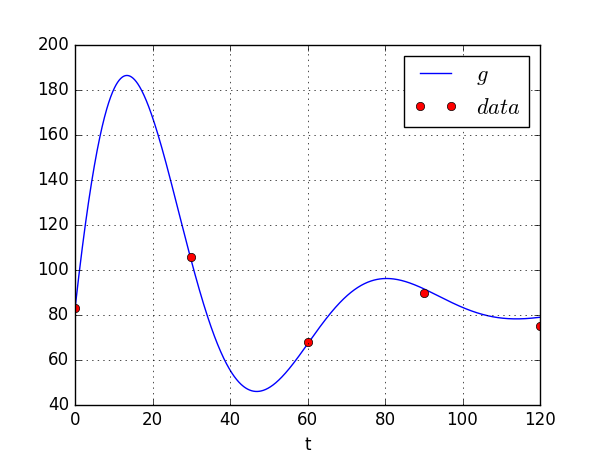} \\
\text{(a) Patient 1 H} & \text{(b) Patient 11 H} \\
\includegraphics[width=2in]{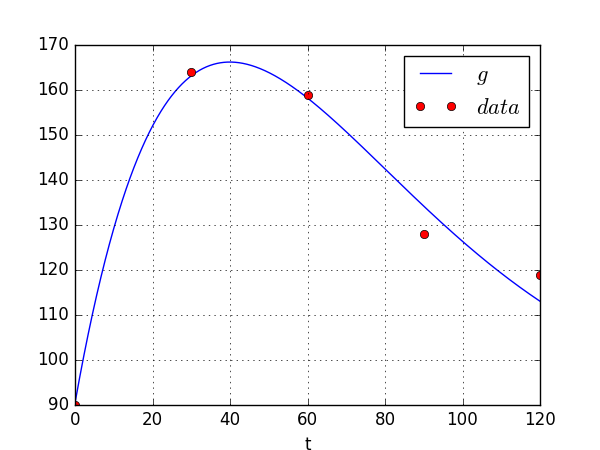} &
\includegraphics[width=2in]{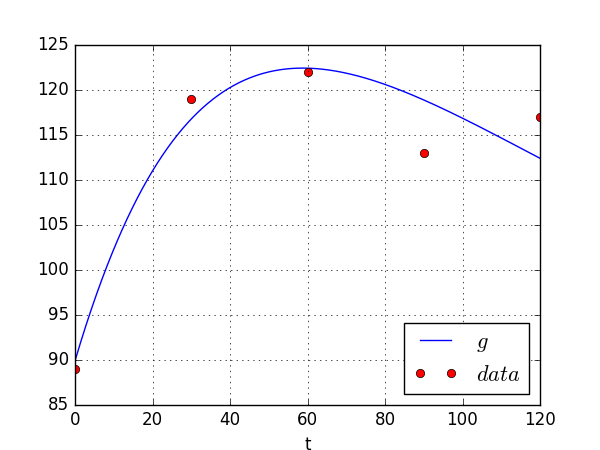} \\
\text{(a) Patient 27 H} & \text{(b) Patient 41 H} \\
 & 
\end{array}
$
\end{center}
\centering{Figure 1. Fitting curves for some healthy patients}
\end{figure}

\begin{figure}[h]
\begin{center}$
\begin{array}{cc}
\includegraphics[width=2in]{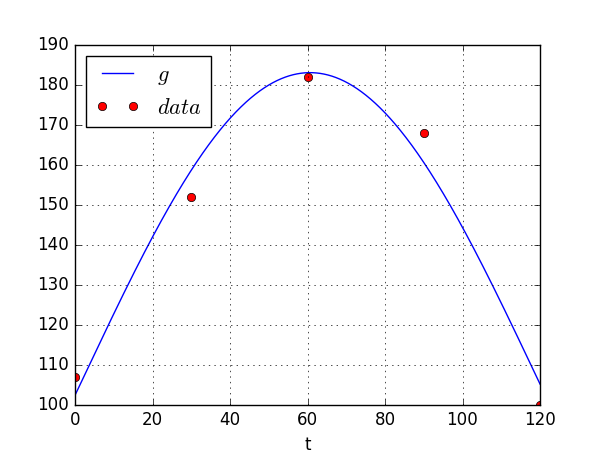} & \includegraphics[width=2in]{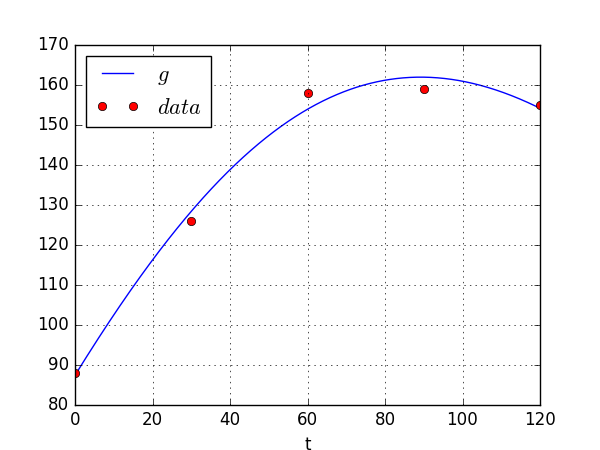} \\
\text{(a) Patient 53 IFG} & \text{(b) Patient 60 IGT} \\
\includegraphics[width=2in]{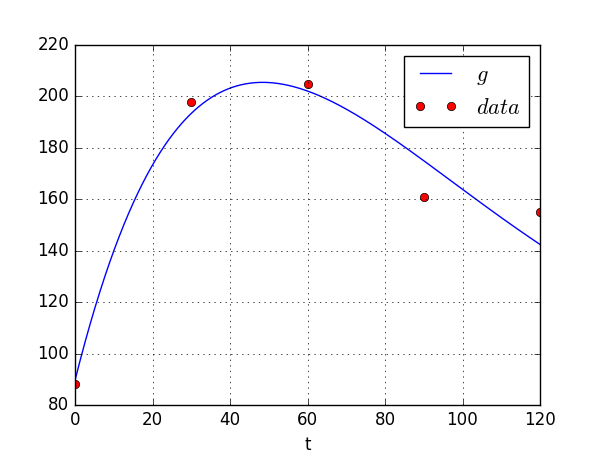} & \includegraphics[width=2in]{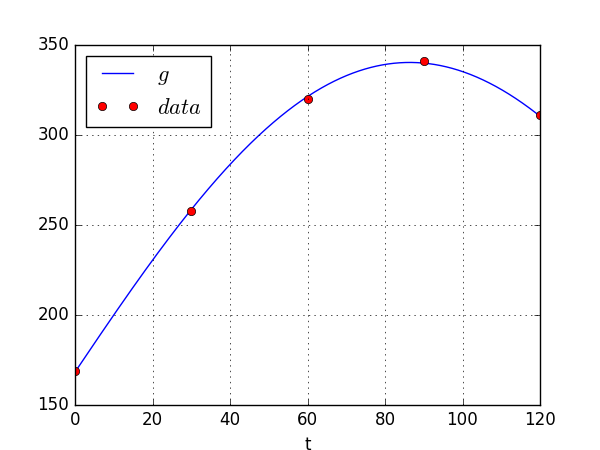}\\
\text{(a) Patient 70 IGT} & \text{(b) Patient 79 T2DM} 
\end{array}
$
\end{center}
\centering{Figure 2. Fitting curves for some ill patients}
\end{figure}

\subsection{Parameters proposal for classification}
 
 Here we use the estimated parameters to classify a patient's condition. The most physically meaningful parameters are 
$A,\alpha$
\begin{itemize}
\item $A$ is related to patient's maximum increase of glucose concentration in response to glucose load.

\item $\alpha$ is related to the patient's ability to attenuate the effect of the glucose load.
\end{itemize}

We regard these as patient's indices to be used for classification of diabetic condition. 

As a first step, we split patients in two groups, healthy patients and patients with a diabetic condition. In the plane $A-\alpha$ we perform a SVM linear separation. 
Here we follow the basic theory, see James et al \cite{JWHT}. 

The classification is successful for 85\% of patients, results in Figure 3. 

\begin{figure}[h] 
   \centering
   \includegraphics[width=4.5in]{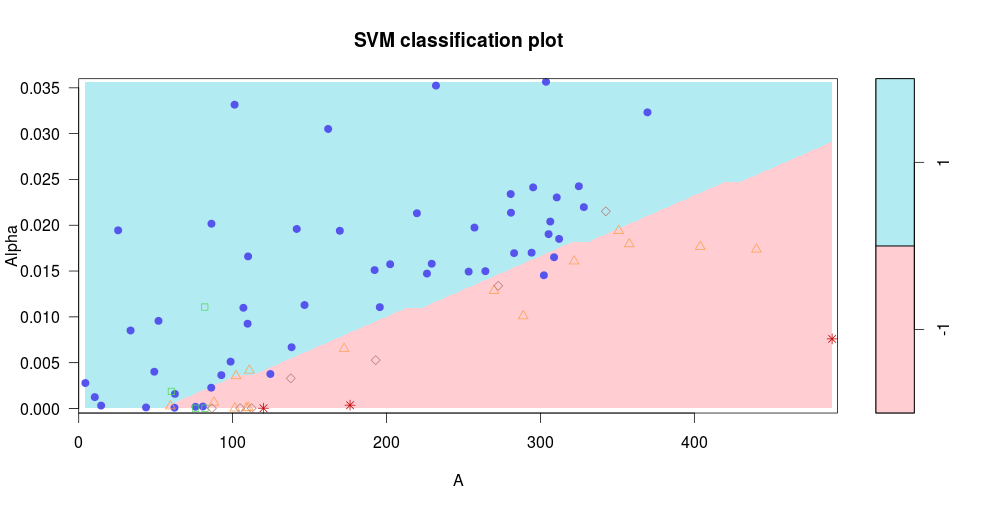} 
   \centering{Figure 3. Healthy ($\bullet$), IFG ($\circ$), IGT ($\vartriangle$), IFG-IGT ($\diamond$).}
\end{figure}

\bigskip

Several remarks are in order.

\begin{itemize}
\item At first sight, one may question the quality of classification in the neighborhood of the separating line. Nevertheless, there is an apparent \emph{clockwise} transition from healthy to T2DM. The latter might have an important implication, namely, early detection of pre-diabetic patients. In such a case, a patient can be controlled with diet and exercise henceforth postponing medication. 

\item These results are obtained using the simplest of models. 

\item This may be regarded a double blind process. Data was gathered independently of the proposed methodology. The latter was tested with the data as provided.

\item More data and planned sampling are needed to confirm our findings.

\end{itemize}

\section{Bayesian Parameter Estimation}

In this section our exposition is deliberatively terse, an excellent first read is Kaipio \& Sommersalo \cite{KaiSom}. For an insightful presentation see Stuart \cite{Stuart}.

In bayesian estimation all variables are random, thus we consider the model
\[
\mathbf{y}=\mathcal{G}(\mathbf{u})+\eta
\]
where $\mathbf{u}$ is the parameter to estimate, $\mathbf{y}$, $\eta$ the noise and $\mathcal{G}$ the observation operator.

An important feature of bayesian estimation is to propose a \emph{prior} probability density function,  $\pi_0$, for the parameter 
$\mathbf{u}$. This prior encompasses all we know about $\mathbf{u}$. In essence it is a modelling problem.

Noise is supposed to be known with density $\rho$ and independent of $\mathbf{u}$. Consequently, the conditional density 
$\pi^{\mathbf{y}}(\mathbf{u})\equiv \pi(\mathbf{u}\vert \mathbf{y})$, the \emph{posterior}, is given from Bayes' formula

\[
\pi^{\mathbf{y}}(\mathbf{u}) \propto \rho(\mathbf{y}-\mathcal{G}(\mathbf{u}))\pi_0(\mathbf{u}).
\]

The point of bayesian estimation is to determine the posterior.  From this, point estimates can be obtained. Namely, the Conditional Mean (CM) and the MAP estimator. For the latter, an optimization problem is to be solved,

\[
\mathbf{u}_{MAP}=argmax\, \rho(\mathbf{y}-\mathcal{G}(\mathbf{u}))\pi_0(\mathbf{u}).
\]

\bigskip

This generalizes a well known deterministic approach. Indeed, consider a  gaussian prior, $\mathbf{u}\sim \mathcal{N}(\mathbf{u}_0,\sigma^2\mathbf{I})$, 

\[
\pi_0(\mathbf{u})\propto exp\left( - \frac{1}{\sigma^2}\left\vert \mathbf{u}-\mathbf{u}_0\right\vert^2 \right)
\]

If noise is also gaussian with zero mean and variance $\gamma$, we have
\[
\mathbf{u}_{MAP}=argmax\left(exp\left( - \frac{1}{\gamma^2}\left\vert y-\mathcal{G}(\mathbf{u})\right\vert^2 \right)
exp\left( - \frac{1}{\sigma^2}\left\vert \mathbf{u}-\mathbf{u}_0\right\vert^2 \right)\right)
\]

\[
\mathbf{u}_{MAP}=argmin\left(\left\vert y-\mathcal{G}(\mathbf{u})\right\vert^2 +
\left(\frac{\gamma}{\sigma}\right)^2\left\vert \mathbf{u}-\mathbf{u}_0\right\vert^2 \right).
\]

Hence, $\mathbf{u}_{MAP}$ is Tikhonov'solution with regularization parameter, $\alpha=\left(\frac{\gamma}{\sigma}\right)^2$.

\bigskip

In our problem

\[
\mathbf{u}=(A,\alpha,\omega,\delta)^t
\]
and assuming gaussian noise we have,
\[
\pi^y(\mathbf{u}) \propto exp\left( - \frac{1}{\gamma^2}\left\vert y-\mathcal{G}(\mathbf{u})\right\vert^2 \right) \pi_0(\mathbf{u})
\]

\bigskip

To sample the posterior we use emcee, an affine invariant Markov Chain Monte Carlo (MCMC) ensemble sampler, Foreman-Mackey et al.
\cite{Foremanetal}.
\bigskip

The prior may influence artificially the determination of the posterior, this is unwanted. In our case we are able to obtain satisfactory results with uninformative (uniform density) priors, only an estimate of the magnitude is required. We use uniform densities:
\begin{itemize}
\item $A\sim U[0.5\, g_m,2.5\, g_M+15]0$.

\item $\alpha \sim U[0,0.1]$.

\item $\omega_0\sim U[0,0.15]$.

\item $\delta\sim U[-2\pi,2\pi]$.
\end{itemize}

Where $g_m, g_M$ are respectively the minimum and maximum of absolute values of $G_0$ shifted glucose concentration data.

\bigskip

Based on experiments, it was observed in Ackerman et al \cite{Ackermanetal} that the parameter $\alpha$ is very sensitive to errors on $G$. Its use was not recommended for a diagnosis criterion. This is certainly true in a deterministic setting. In our approach we allow large observation errors with a gaussian model with standard deviation $\gamma=5$, a typical value in the literature. An advantage of the bayesian approach, is that uncertainty of our point estimates  is readily quantified by means of the marginal densities of the posterior. We found that densities for $\alpha$ are highly concentrated at the point estimates, mostly unimodal. For the patients above see Figures 4 and 5. Hence, we argue that the estimates are reliable.

\begin{figure}[h]
\begin{center}$
\begin{array}{cc}
\includegraphics[width=40mm]{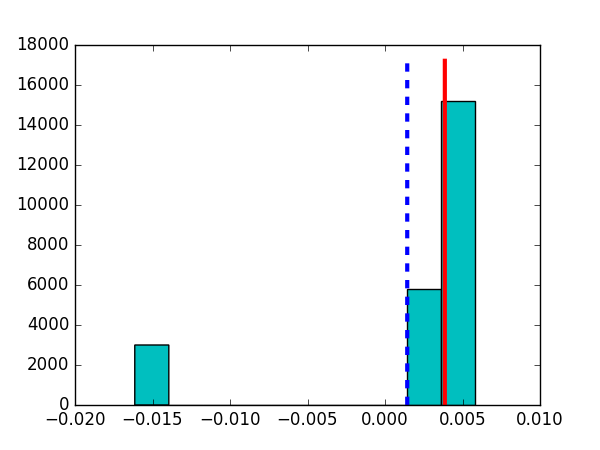} & \includegraphics[width=40mm]{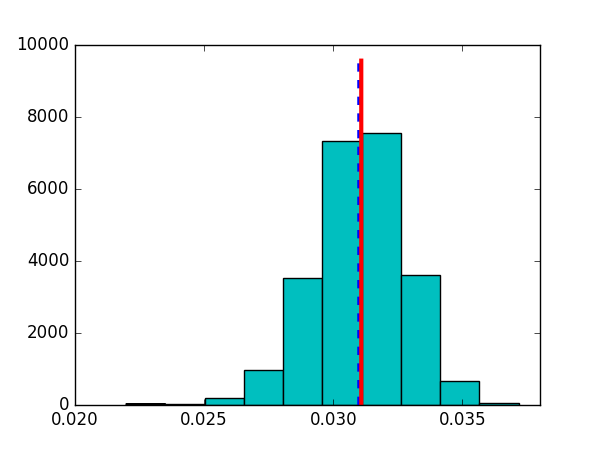} \\
\text{(a) Patient 1 IFG} & \text{(b) Patient 11 IGT} \\
\includegraphics[width=40mm]{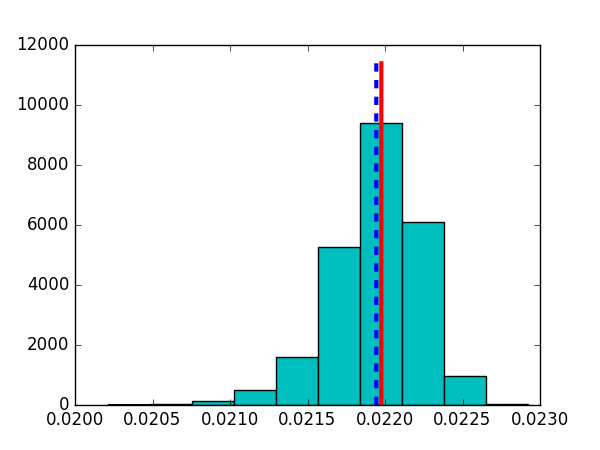} & \includegraphics[width=40mm]{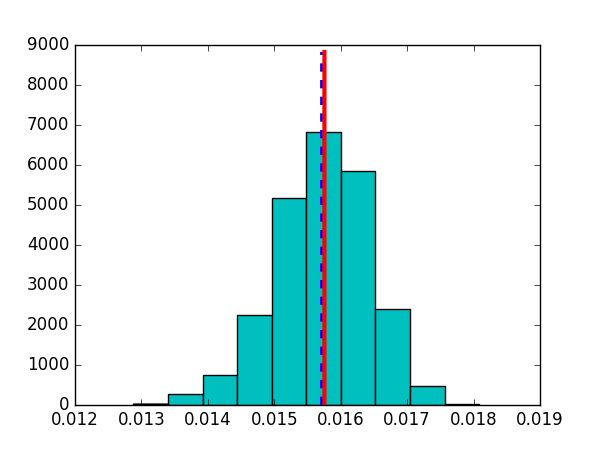} \\
\text{(a) Patient 27 IGT} & \text{(b) Patient 41 T2DM} 
\end{array}$
\end{center}
\centering{Figure 4. Marginal posterior densities for healthy patients. MAP solid line. CM dashed line.}
\end{figure}

\begin{figure}[h]
\begin{center}$
\begin{array}{cc}
\includegraphics[width=40mm]{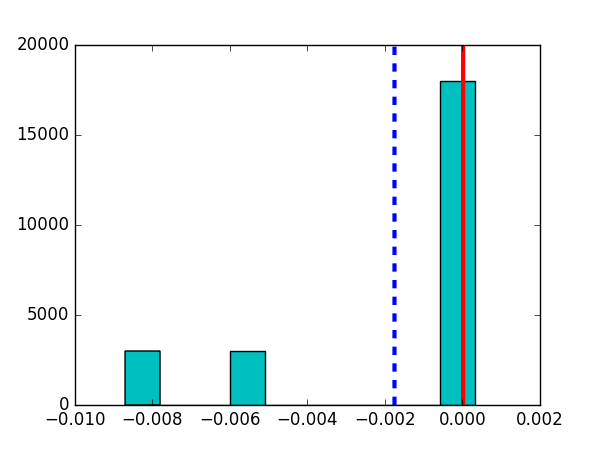} & \includegraphics[width=40mm]{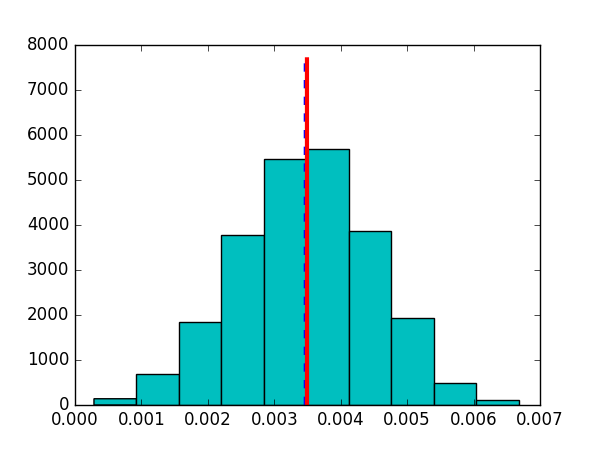} \\
\text{(a) Patient 53 IFG} & \text{(b) Patient 60 IGT} \\
\includegraphics[width=40mm]{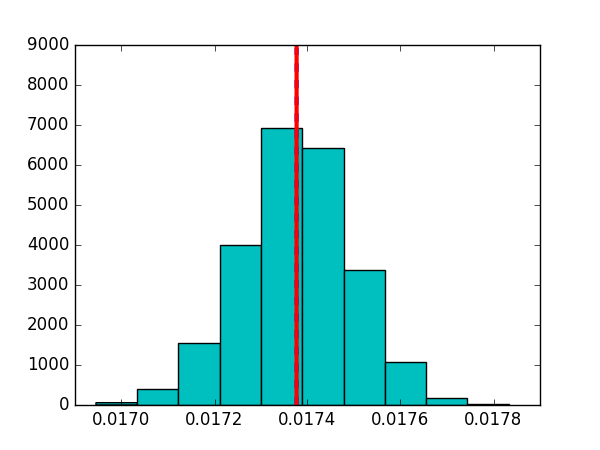} & \includegraphics[width=40mm]{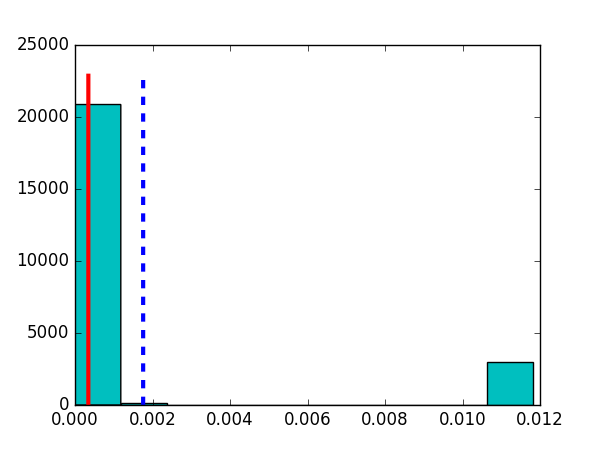} \\
\text{(a) Patient 70 IGT} & \text{(b) Patient 79 T2DM} 
\end{array}$
\end{center}
\centering{Figure 5. Marginal posterior densities for ill patients. MAP solid line. CM dashed line.}
\end{figure}

\section{Conclusions and future work}

In this work we have considered a glucose-insulin interaction model for parameter estimation. The parameters are obtained from bayesian estimation constrained to a simple ordinary differential equation. Two of these parameters are chosen as patient's indices. A SVM classification algorithm show the potential of these indices to  determine a patient's healthy or diabetic condition.  

As a proof of concept for the introduced methodology, we started with the simplest of models obtaining satisfactory results.  Consideration of more sophisticated glucose-insulin models is part of our current and future work. Also, more data is required to be conclusive.

\bigskip
\center{\textbf{Acknowledgements}}

M. A. Moreles thanks the support of ECOS-NORD through the project: 000000000263116/M15M01.

\end{document}